\documentstyle[12pt,aps,prc,epsfig,preprint]{revtex}
\tightenlines
\begin{document}

\begin{titlepage}
\title{\vspace*{10mm}\bf
\Large Evidence for partial chiral symmetry restoration from pionic atoms}
\vspace{6pt}

\author{  E.~Friedman  \\
{\it Racah Institute of Physics, The Hebrew University, Jerusalem 91904,
Israel\\}}

\vspace{4pt}
\maketitle

\begin{abstract}

Extensive  data on strong interaction effects in pionic atoms
are analyzed with a density-dependent isovector scattering amplitude
suggested recently by Weise to result from a density dependence of the
pion decay constant. Most of the so-called `missing $s$-wave repulsion'
is removed when adopting this approach, thus indicating
a partial chiral symmetry restoration in dense matter.
The resulting potentials describe
quite well also elastic scattering of 20 MeV pions on Ca. Further tests
with elastic scattering are desirable.
\newline \newline
$PACS$: 13.75.Gx; 25.80Dj
\newline
\newline \newline
Corresponding author: E. Friedman, \newline
Tel: +972 2 658 4667,
FAX: +972 2 658 6347, \newline
E mail: elifried@vms.huji.ac.il

\end{abstract}

\centerline{\today}
\end{titlepage}

The interaction of low energy pions with nuclei has been known for years
to be described well by  a theoretically-motivated 
phenomenological optical
potential \cite{EEr66}, particularly at 
zero energy where  strong interaction
effects in pionic atoms have been  studied extensively both experimentally
and theoretically
 \cite{BFG97}. The traditional method of spectroscopy of pionic X-rays
has been supplemented very recently by the observation of `deeply bound'
pionic atom states through the $(d,^3$He) 
reaction \cite{YHI96,GGK00}, thus adding a
new dimension to the ability to study pion interactions
at threshold in the nuclear medium.
Whereas the $p$-wave part of the pion-nucleus optical potential,
which is effective only near the nuclear surface,
is described rather well by the free pion-nucleon amplitudes (plus a two-nucleon
absorption term), this is not the case for the $s$-wave part of the potential.
This part of the potential, which is effective throughout the nuclear volume,
is a natural source of information on possible modifications by the nuclear
medium of the pion-nucleon interaction. This is the topic
of the present Letter which deals with the strong $s$-wave repulsion of
pions in nuclear matter and its possible origins in a density dependence of the
pion decay constant which reflects the change of QCD vacuum structure
in dense matter.

The interaction between low energy pions and  nuclei is traditionally
described \cite{BFG97} by an optical potential as follows:

\begin{equation} \label{EE1}
2\mu V_{{\rm opt} }(r) = q(r) + \vec \nabla \cdot \alpha(r) \vec \nabla
\end{equation}

\noindent
with the $s$-wave part given by

\begin{eqnarray} \label{EE1s}
q(r) & = & -4\pi(1+\frac{\mu}{M})\{\bar b_0[\rho_n(r)+\rho_p(r)]
  +b_1[\rho_n(r)-\rho_p(r)] \} \nonumber \\
 & &  -4\pi(1+\frac{\mu}{2M})4B_0\rho_n(r) \rho_p(r),
\end{eqnarray}

\noindent
where $\rho_n$ and $\rho_p$ are the neutron and proton density
distributions normalized to the number of neutrons $N$ and number
of protons $Z$, respectively, $\mu$ is the pion-nucleus reduced mass
and $M$ is the mass of the nucleon. 
The parameter $\bar b_0$ is given in terms of the 
pion-nucleon (minus) isoscalar
and isovector scattering lengths $b_0$ and $b_1$, respectively,

\begin{equation} \label{b0b}
\bar b_0 = b_0 - \frac{3}{2\pi}(b_0^2+2b_1^2)k_F,
\end{equation}
where $k_F$ is the local Fermi momentum. This second order term is included
because of the extremely small value of $b_0$ 
\cite{SBG01} and it will be shown to play 
a decisive role in what follows.
The term with the complex
parameter $B_0$ represents
absorption on a neutron-proton pair.
The term with $\alpha(r)$ is referred to
as the $p$-wave potential, see Eqs.(20-22) of \cite{BFG97}.

Values of the various parameters of the potential are obtained from fits
to experimentally determined strong interaction level shifts and widths
and `upper' level yields. Modern data sets containing at least 50 data
points along the periodic table lead to rather well defined values for the 
various parameters and to good agreement between calculation and
experiment, with typically
$\chi ^2$ per point of about 2.
 Addressing the real part of the $s$-wave potential,
it has been found \cite{BFG97}
that both $b_0$ and Re$B_0$ are well determined by the
data, contrary to earlier conclusions \cite{SMa83} which were based on
considerably more restricted data.  
A somewhat confusing situation arises
when values of Re$B_0$ are  found to be large and repulsive, i.e. 
3 to 5 times larger
than the values of Im$B_0$, whereas expectations are that Re$B_0$ is attractive
and of about the same magnitude as the imaginary part. This unexpected
phenomenological repulsion has been referred to as a `missing 
repulsion' \cite{BFG97,EGR95}.

\begin{figure}
\epsfig{file=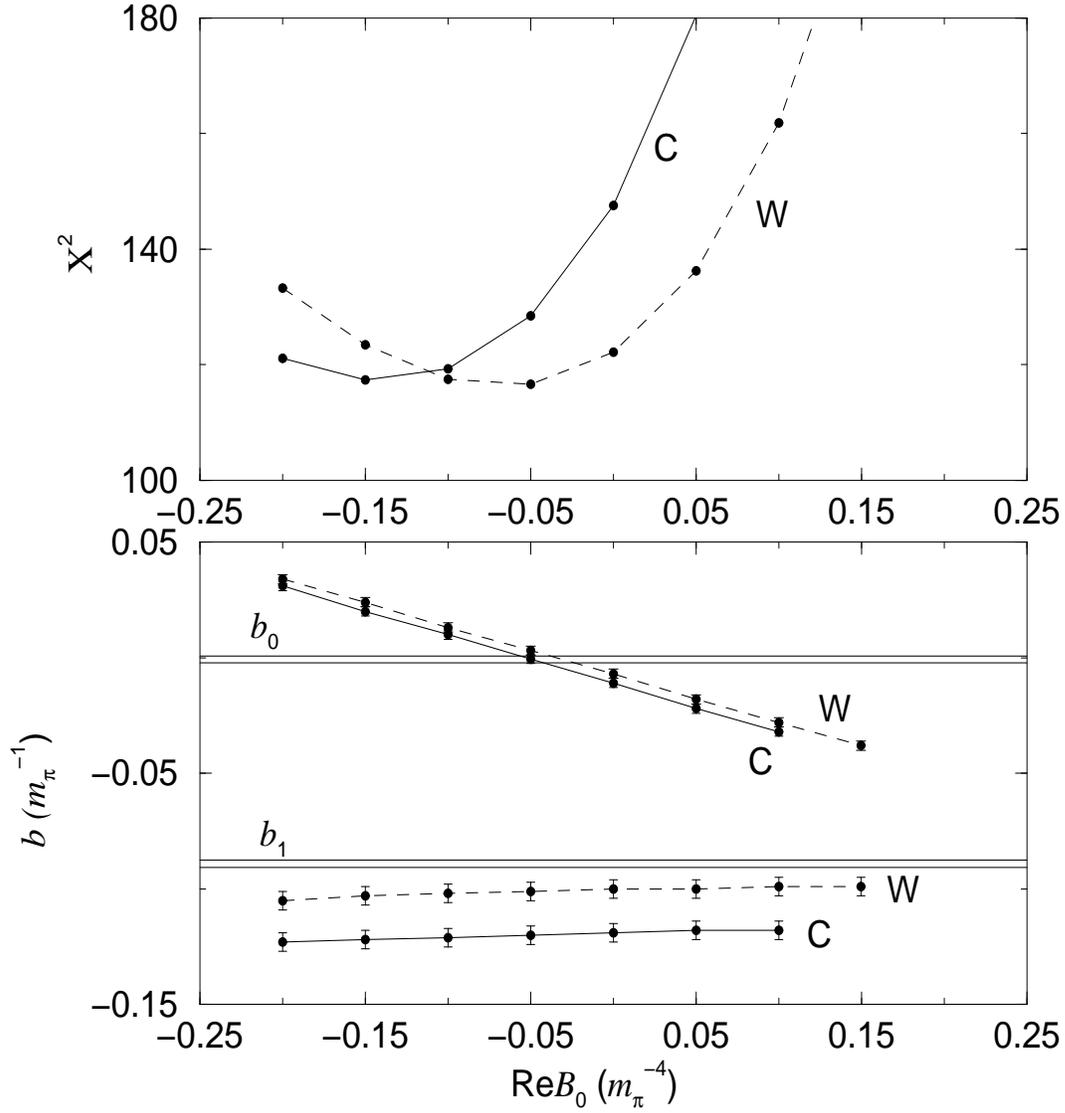, height=150mm,width=140mm}
\caption{Fits to pionic atom data as function of Re$B_0$. Upper part: values
of $\chi ^2$, lower part: values of $b_0$ and $b_1$. Also shown as horizontal
bands are values of $b_0$ and $b_1$ for the free pion-nucleon interaction.
C stand for the conventional potential, W stand for the Weise prescription
(Eq.(\ref{ddb1}))}.
\label{fig:CW}
\end{figure}

In the present work fits have been made to 60 experimental values of level
shifts, widths and upper level yields for targets from $^{16}$O to U, 
including the very recently determined binding energies and widths 
for the deeply
bound 1s and 2p states in $^{205}$Pb \cite{Gil01,Gei01}. 
As the parameters of the
linear term of the $p$-wave part of the potential were always found to 
be very close to the free pion-nucleon values when the Lorentz-Lorenz parameter
$\xi$ was close to 1, we have kept subsequently these parameters fixed at the
free pion-nucleon values together with $\xi$=1 and varied only the parameters
of the $s$-wave part of the potential and the phenomenological quadratic
(absorptive) term in the $p$-wave part.  The  latter, traditionally denoted
by $C_0$, was found to be independent
of variations in the $s$-wave  part of the potential and its real part was
essentially  zero.
In order to focus on the various components of the real part of the $s$-wave
potential, we show in
Fig. \ref{fig:CW}  results of such fits, as a function of the parameter
Re$B_0$. The upper part shows values of $\chi ^2$ for the 60 data points
and the lower part shows the corresponding values of the other  parameters
of the real part
of the $s$-wave potential, namely, $b_0$ and $b_1$.  The values of Im$B_0$ were 
found to be remarkably constant at  0.056 $m_{\pi}^{-4}$.
Also shown as horizontal bands in the lower part are the free
pion-nucleon values of $b_0$ and $b_1$ which have been determined 
very recently to high precision \cite{SBG01}.

Three features are easy to
observe from the solid curves labelled  `C': 
(i) the values of all three parameters are well
determined, (ii) the parameter Re$B_0$ is repulsive and large 
(Re$B_0 \approx -3$Im$B_0$) and (iii) $b_0$ and $b_1$ are well determined and
are significantly different from the corresponding free pion-nucleon values. 
Note that the
parameter $b_1$ is found to be more than 35\% larger 
in absolute value than its free pion-nucleon
value and  that it contributes significantly to the repulsion, also in $N=Z$
nuclei, through the $b_1^2$ term in Eq.(\ref{b0b}).  In fact, for $^{40}$Ca
it contributes as much as 35\% of the real potential. Thus 
the `missing repulsion'
appears as a very repulsive dispersion term Re$B_0$ {\it and} an
enhanced $b_1$ parameter. 
This information is lost when one adopts the `effective density' approach
of lumping together $b_0$ and Re$B_0$ \cite{SMa83,SHO95}.
As the free pion-nucleon $b_0$ is extremely small and the empirical
values of $b_0$ are much smaller than values of $b_1$,  we discuss only the 
$b_1$ parameter, where
the extra repulsion observed 
 may  be associated with medium
modification of the pion-nucleon interaction.

The in-medium $s$-wave interactions of pions have been discussed very
recently by Weise \cite{Wei01} 
in terms of partial restoration of chiral symmetry in dense matter
where the isospin-odd in-medium pion-nucleon amplitude
is inversely proportional to the square of the 
 pion decay constant $f_\pi$. The square
of the latter is given, in leading order, 
 as a linear function of the nuclear density,

\begin{equation} \label{eq:fpi2}
f_\pi ^{*2} = f_\pi ^2 - \frac{\sigma _N}{m_\pi ^2} \rho
\end{equation}
with $\sigma_N$ the pion-nucleon sigma term.
This leads to a density-dependent isovector amplitude such that $b_1$ becomes

\begin{equation}\label{ddb1}
b_1(\rho) = \frac{b_1(0)}{1-2.3\rho}
\end{equation}
for $\sigma _N$=50 MeV and
with $\rho$ in units of fm$^{-3}$. 
Note that expanding this expression in powers of the density leads naturally
to a repulsive $\rho ^2$ term in the pion-nucleus potential.
We have introduced this expression for
$b_1$ into the potential, using for $\rho$ the local nuclear density 
$\rho (r)$ and repeated
the fits to the experimental results. 
The dashed curves labelled `W' in Fig. \ref{fig:CW} show  the results
obtained with this prescription.  
From the minimum of $\chi^2$ it is clear that now Re$B_0$ is 
considerably less repulsive (Re$B_0 \approx -$Im$B_0$) and that $b_0$ agrees
with the free pion-nucleon value. The value of $b_1$ is now much closer
to the free pion-nucleon value, although still a little more repulsive. 
It is seen, therefore, that the introduction of the theoretically motivated
medium dependence into the repulsive
 terms containing $b_1$ removes a major fraction of the 
excessive phenomenological repulsion, as evidenced by the significantly reduced
magnitude of both $b_1$ and Re$B_0$.
The best fit parameters for the above potentials are summarized in the
first two rows of Table \ref{tab:para}. The third row (`W65') is for the same
prescription but with a larger value for the pion-nucleon sigma term of
$\sigma _N$=65 MeV (not shown in the figure.) It is seen 
from the table that for this value
of $\sigma _N$ the empirical $b_0$ and  $b_1$ parameters 
are consistent with the free
pion-nucleon values.
Higher order terms have also been considered very recently \cite{KWe01}
and found to be small.

\begin{figure}
\epsfig{file=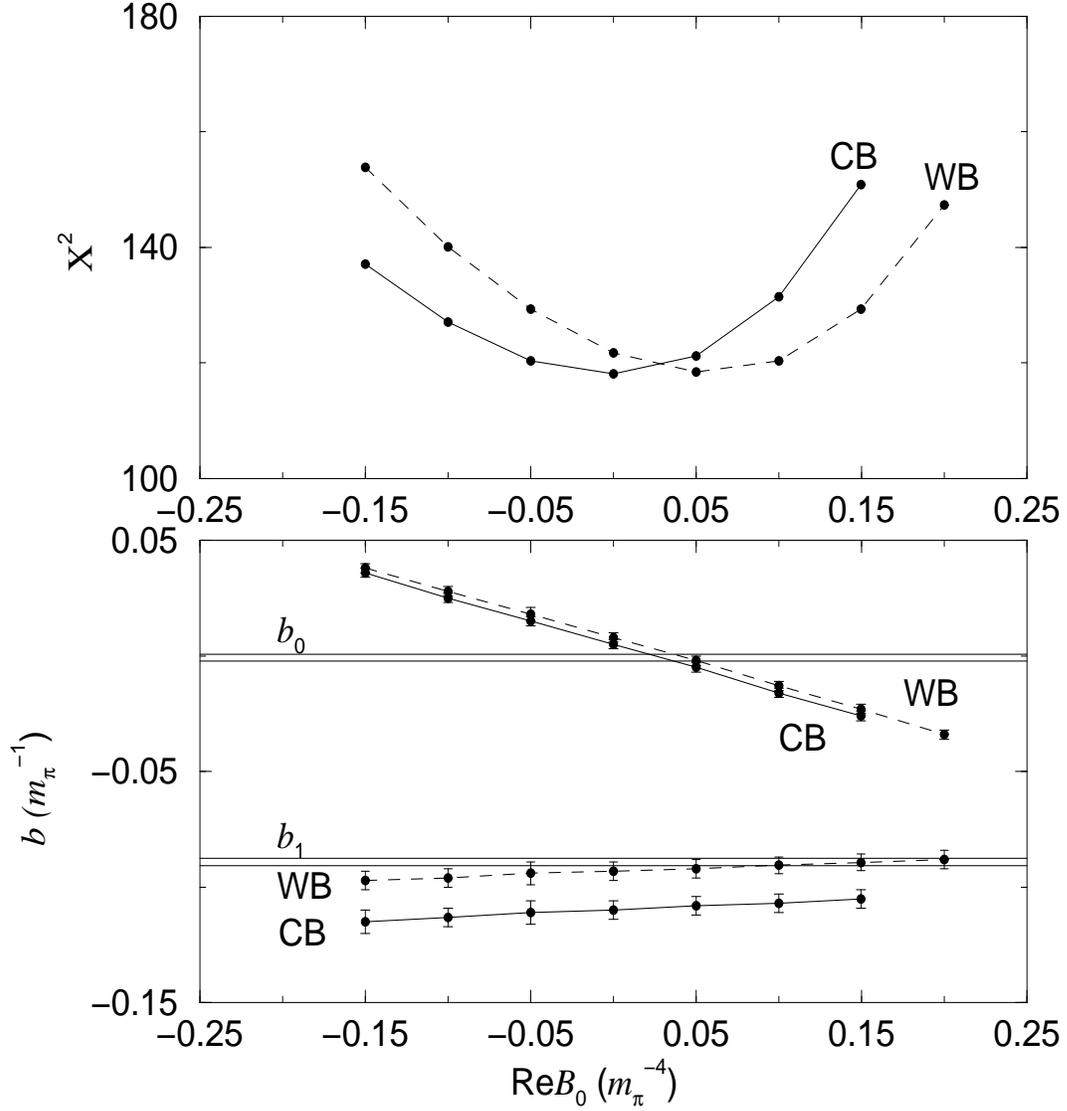, height=150mm,width=140mm}
\caption{Same as Fig.\ref{fig:CW} but with the RIA term included. CB stand
for the conventional potential, WB stand for the Weise prescription.}
\label{fig:CWB}
\end{figure}

Among several previous attempts to account for the missing $s$-wave repulsion
we mention a relativistic impulse approximation (RIA) approach \cite{GJF92}
which showed, following Birbrair and others 
\cite{BFK83,GLM91,BGr91}, that  a specific version of the RIA 
is able to provide a significant part of the missing repulsion
through the modification of the {\it nucleon} mass in the nuclear medium.
We have therefore
looked again into this possibility, noting, however, that in Ref.\cite{GJF92}
it was shown that there was {\it no unique} way of introducing RIA effects
into the pion-nucleus interaction \cite{CJe94}.
This specific version of the RIA 
 was included
using the following parameterization \cite{BGr91}

\begin{equation}
\frac{M(\rho )}{M(0)} = \frac{1}{1 + a \rho} 
\end{equation}
with $a$=2.7 fm$^3$, amounting to $M(\rho)/M(0)=0.7$ for 
the nucleon mass ratio at normal nuclear
density. The results are shown in Fig.\ref{fig:CWB} where
 it is seen that when the RIA correction is
applied to the conventional potential (`CB', solid curves) no repulsion
is required through the $B_0$ term, but the values of $b_0$ and $b_1$ are still
not in agreement with the free pion-nucleon values. 
Also seen from the figure is that when the RIA term is included together
with the 
theoretically motivated density dependence 
of $b_1$ Eq.(\ref{ddb1})  (`WB' dashed
curves), 
an {\it attractive}
Re$B_0$ is found whose magnitude is close to the magnitude of the absorptive
part and, then, both $b_0$ and $b_1$ agree with the 
corresponding free pion-nucleon values.
The best fit values of the potential parameters for these two versions
of the potential (with $\sigma _N$=50 MeV) are 
also summarized in Table \ref{tab:para}.

The value of the real potential
at the center of the $^{208}$Pb nucleus has received some attention
recently \cite{WBW97,FGa98}. All five potentials yield values between
34 and 39 MeV for this quantity. Obviously these values are extrapolated
from the better determined values of the potential near the nuclear surface.
Taking e.g. the real potential at the 50\% density point, then all five
potentials yield  values between 12.4 and 13.7 MeV for this quantity.

\begin{figure}
\epsfig{file=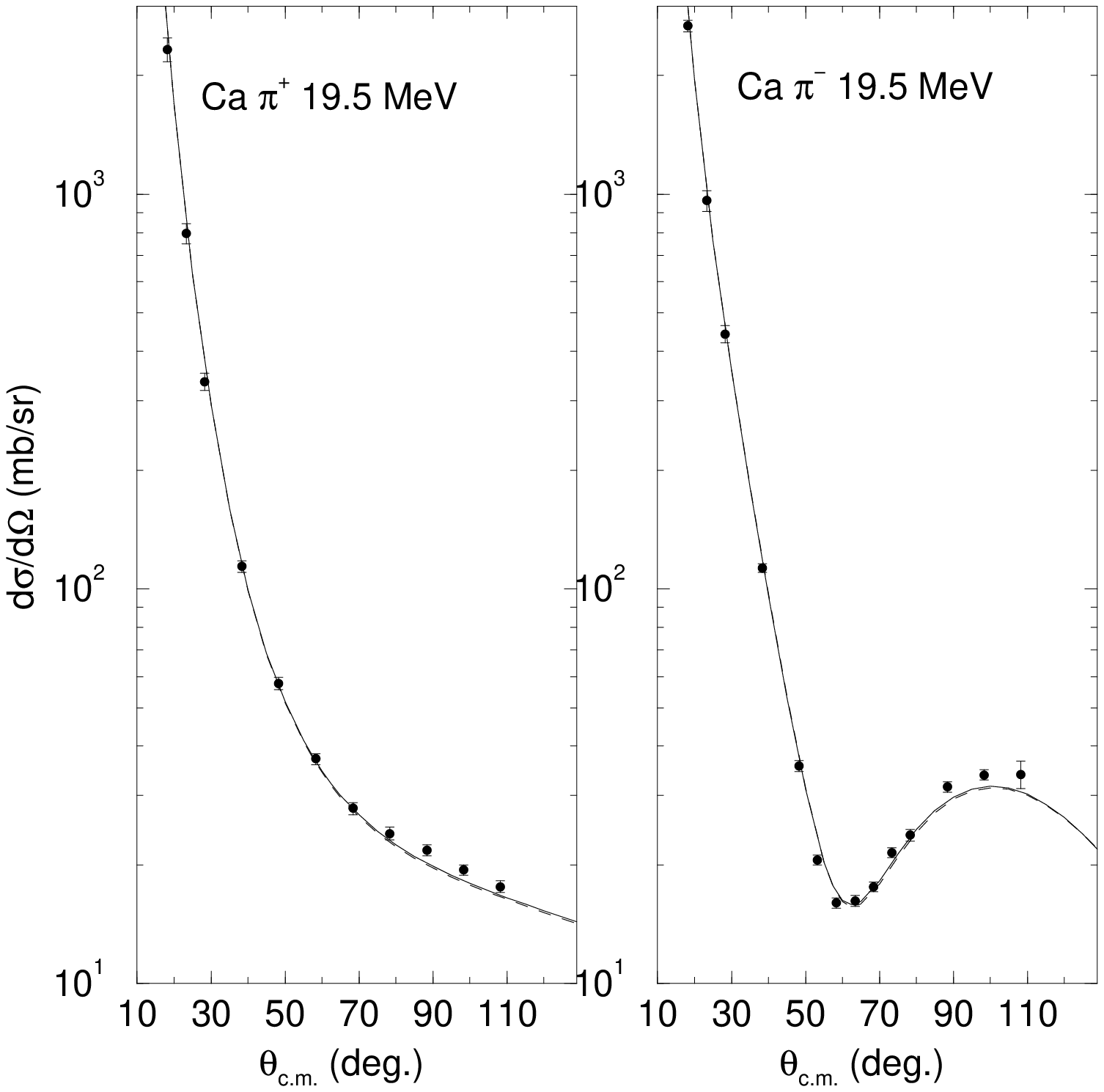, height=120mm,width=140mm}
\caption{Elastic scattering of 19.5 MeV pions by Ca. Experimental results 
from \protect \cite{Wri88}.
Solid lines: potential CB of Table \ref{tab:para}, dashed lines: potential
WB of the table.}
\label{fig:scatt}
\end{figure}

It is interesting to study the above mentioned features at energies just above
threshold through the elastic scattering of  low
energy pions by nuclei, thus testing further the validity of the chirally
motivated approach. 
Indeed it has been shown 
\cite{SMa83,SMC79,MFJ89} that  pion-nucleus
potentials  develop smoothly from the bound states regime to the 
elastic scattering regime.
Here we examine only  the elastic scattering of 19.5 MeV pions by Ca
with the help of the 
experimental results of Wright et al. \cite{Wri88} which seem to be the
only fairly extensive data for $\pi ^+$ and $\pi^-$ on the same nucleus
and from the same experiment
at such low energies. Using the parameters of Table \ref{tab:para} we have
calculated the differential cross sections for elastic scattering of pions
by Ca at 19.5 MeV and found reasonable agreement with the data. The agreement
with the data is a little better for the two potentials that include the RIA
corrections (last two rows of the table). 
Fig. \ref{fig:scatt} shows comparisons between
experiment and calculations for  these two potentials.
Improving the agreement
by adjusting the complex parameter $B_0$, we find that the values of
 Im$B_0$ hardly change 
at all but Re$B_0$ has to be made a little more repulsive,
typically by 0.02 $m_\pi ^{-4}$. Alternatively, if $b_0$ is adjusted, the
extra repulsion is compatible with the energy dependence of this parameter.
It is therefore concluded that the limited experimental results for elastic
scattering of very low energy pions by nuclei support the picture that emerges
from the extensive studies of pionic atoms regarding the nature of the missing
$s$-wave repulsion. However, the data on elastic scattering are for one nucleus only, whereas we have included 23 nuclei in the study of pionic atoms. Therefore
 precision data for the elastic scattering of very low energy $\pi ^{\pm}$ 
on some additional nuclei are
desirable.

In conclusion, we have shown that most of the `missing $s$-wave repulsion'
in the interaction of pions at threshold with nuclei can be removed by
adopting a density-dependent isovector amplitude as suggested by Weise
\cite{Wei01} to result from a density dependence of the pion decay constant.
The underlying picture is that of partial restoration of chiral symmetry
in dense matter.
When an additional RIA term is included, the best fit pionic atom potential
is in full agreement with the chirally motivated model based on the 
{\it free} pion-nucleon amplitudes.

\vspace{8mm}
 
I wish to acknowledge fruitful discussions with H. Clement, A. Gal and 
G.J. Wagner.
This research was partly supported by the trilateral DFG contract
GR 243/51-2.

\begin{table}
\caption{Parameter values from fits to 60 pionic atom data points. Other
$p$-wave parameters were held fixed at $c_0$=0.22$m_\pi^{-3}$,
$c_1$=0.18$m_\pi^{-3}$ and $\xi$=1. The free pion-nucleon values 
\protect \cite{SBG01}
are $b_0=-0.0001^{+0.0009}_{-0.0021} m_\pi ^{-1}$ and
$b_1=-0.0885^{+0.0010}_{-0.0021} m_\pi ^{-1}$}
\label{tab:para}
\begin{tabular}{lcccccc}
potential&$\chi ^2$&$b_0$ ($m_\pi^{-1}$)&$b_1$ ($m_\pi^{-1}$)&
 Re$B_0$ ($m_\pi^{-4}$)& Im$B_0$ ($m_\pi^{-4}$)&Im$C_0$ ($m_\pi^{-6}$) \\
\hline
C & 117.3 & 0.018$\pm$0.010&$-$0.122$\pm$0.004&$-$0.14$\pm$0.04&
0.056$\pm$0.002&
0.056$\pm$0.004 \\
W & 116.3 & 0.007$\pm$0.009&$-$0.102$\pm$0.004&$-$0.06$\pm$0.04&0.056$\pm$0.002&
0.056$\pm$0.004 \\
W65 & 117.7 & 0.001$\pm$0.009&$-$0.095$\pm$0.004&$-$0.03$\pm$0.04&0.055$\pm$0.002&0.055$\pm$0.004 \\
CB & 118.2 & 0.005$\pm$0.010&$-$0.110$\pm$0.004&0.00$\pm$0.04&0.056$\pm$0.002&
0.056$\pm$0.004 \\
WB & 118.3 &$-$0.004$\pm$0.010&$-$0.092$\pm$0.004&0.06$\pm$0.04&0.055$\pm$0.002&
0.056$\pm$0.004 \\
\end{tabular}
\end{table}

\begin{thebibliography}{bat96}

\bibitem{EEr66}M. Ericson, T.E.O. Ericson, Ann. Phys.  [NY] 36 (1966) 323.

\bibitem{BFG97}For a recent review see  C.J. Batty, E. Friedman, A. Gal, 
Phys. Rep. 287
        (1997) 385.

\bibitem{YHI96}T. Yamazaki et al., 
Z. Phys. A 355 (1996) 219.

\bibitem{GGK00}H. Gilg et al., 
Phys. Rev. C 62 (2000) 025201;  K. Itahashi et al., Phys. Rev. C 62 (2000) 
025202.

\bibitem{SBG01}H.-Ch. Schr\"oder et al., ETHZ-IPP PR-2001-1 preprint;
Euro. Phys. J. in press; H.-Ch. Schr\"oder et al., Phys. Lett. B 469
(1999) 25.


\bibitem{SMa83}R. Seki, K. Masutani, Phys. Rev. C 27 (1983) 2799.

\bibitem{EGR95}E. Oset, C. Garc\'{\i}a-Recio, J. Nieves, Nucl.
Phys. A 584 (1995) 653.


\bibitem{Gil01}A. Gillitzer, Proc. Int. Workshop XXIX on Gross Properties 
of Nuclei and
Nuclear Excitations, Hirschegg, Austria, Jan 14-20, 2001, p. 56.

\bibitem{Gei01}H. Geissel et al., submitted to Phys. Rev. Lett. (2001).


\bibitem{SHO95}L.L. Salcedo, K. Holinde, E. Oset, C. Sch\"utz,
Phys. Lett. B 353 (1995) 1.

\bibitem{Wei01}W. Weise, Nucl. Phys. A 690 (2001) 98.

\bibitem{KWe01}N. Kaiser, W. Weise, Phys. Lett. B 512 (2001) 283.

\bibitem{GJF92}A. Gal, B.K. Jennings, E. Friedman,
Phys. Lett. B 281 (1992) 11.

\bibitem{BFK83}B.L. Birbrair, V.N. Fomenko, A.B. Gridnev,
Yu.A. Kalashnikov,
J. Phys. G: Nucl. Phys. 9 (1983) 1473; 11 (1985) 471.

\bibitem{GLM91}P.F.A. Goudsmit, H.J. Leisi, E. Matsinos,
Phys. Lett. B 271 (1991) 290.

\bibitem{BGr91}B.L. Birbrair, A.B. Gridnev, Nucl. Phys. A 528 (1991) 647; B.L.
Birbrair, A.B. Gridnev, L.P. Lapina, A.A. Petrunin, A.I. Smirnov, Nucl.
Phys. A 547 (1992) 645.

\bibitem{CJe94}S. Chakravarti, B.K. Jennings, Phys. Lett. B 323 (1994) 253.

\bibitem{WBW97}T.Waas, R. Brockmann, W. Weise, Phys. Lett. B 405 (1997) 215.

\bibitem{FGa98}E. Friedman, A. Gal, Phys. Lett. B 432 (1998) 235.


\bibitem{SMC79}K. Stricker, H. McManus, J.A. Carr, Phys. Rev. C 19
(1979) 929.

\bibitem{MFJ89}O. Meirav et al., 
 Phys. Rev. C 40 (1989) 843.

\bibitem{Wri88}D.H. Wright et al., Phys. Rev. C 37 (1988) 1155.


\end{thebibliography}
\end{document}